\documentclass[aps,prl,floatfix,twocolumn,superscriptaddress,preprintnumbers,nofootinbib]{revtex4-1}

\usepackage{textcomp} 
\usepackage{slashed}
\usepackage{epsfig,latexsym,cancel,amssymb,amsmath,verbatim,mathrsfs}
\usepackage{color}
\usepackage{graphicx}
\usepackage{enumitem}
\usepackage[colorlinks,citecolor=blue]{hyperref}

\newcommand{\beq}{\begin{equation}}
\newcommand{\eeq}{\end{equation}}
\newcommand{\bea}{\begin{eqnarray}}
\newcommand{\eea}{\end{eqnarray}}

\newcommand{\nn}{\nonumber}
\newcommand{\tttt}{t\bar{t}t\bar{t}}

\allowdisplaybreaks[4]

\begin{document}

\title{Probing Top-philic New Physics via Four-Top-Quark Production}

\author{Qing-Hong Cao}
\email{qinghongcao@pku.edu.cn}
\affiliation{Department of Physics and State Key Laboratory 
of Nuclear Physics and Technology, Peking University, Beijing 100871, China}
\affiliation{Collaborative Innovation Center of Quantum Matter, Beijing 100871, China}
\affiliation{Center for High Energy Physics, Peking University, Beijing 100871, China}

\author{Jun-Ning Fu}
\email{fujunning@pku.edu.cn}
\affiliation{Department of Physics and State Key Laboratory 
of Nuclear Physics and Technology, Peking University, Beijing 100871, China}

\author{Yandong Liu}
\email{ydliu@bnu.edu.cn}
\affiliation{Key Laboratory of Beam Technology of Ministry of Education, College of Nuclear Science and Technology, Beijing Normal University, Beijing 100875, China}
\affiliation{Beijing Radiation Center, Beijing 100875, China}

\author{Xiao-Hu Wang}
\email{xiaohuwang@pku.edu.cn}
\affiliation{Department of Physics and State Key Laboratory 
of Nuclear Physics and Technology, Peking University, Beijing 100871, China}

\author{Rui Zhang}
\email{zhangr@ihep.ac.cn}
\affiliation{Theoretical Physics Division, Institute of High Energy Physics, Beijing 100049, China}

\begin{abstract}
We explore constraints on various new physics resonances from four top-quark production based on current experimental data. Both light and heavy resonances are studied in the work. A comparison of full width effect and narrow width approximation is also made. 

\end{abstract}

\maketitle

\noindent\textbf{1. Introduction.}

Despite its rare rate four top-quark ($t\bar{t}t\bar{t}$) production in hadron collision was discussed even before the discovery of the top quark~\cite{Barger:1991vn}. The four-top channel is very special in the Standard Model (SM) as it involves both the quantum chromodynamics (QCD) and electroweak (EW) interactions and the strengths of the both  interactions are comparable. In particular,  thanks to the heavy top-quark mass, the Yukawa interaction between the Higgs boson and the top quark is fairly large such that the leading electroweak contribution from the Higgs-top interaction is as important as the one from the QCD interaction. These two interactions are interwoven more when one calculates the higher order quantum corrections; as a result, the theoretical prediction of the four-top production cross section is sensitive to the choice of renormalization scale which demands two or more loop calculation to reduce the theoretical uncertainties~\cite{Frederix:2017wme}. 
In addition the four-top channel involves complicated kinematics which enables the interference between the QCD diagrams and heavy EW resonances to yield a sizable contribution. Therefore, the four-top production is a good parade ground to test new physics (NP) beyond the SM. 

In addition, the four-top channel serves well for probing both the magnitude and CP phase of the top-Higgs interaction without assumptions on the decay of the Higgs boson, e.g. neither the branch ratio of particular decay mode nor the total decay width of the Higgs boson~\cite{Cao:2016wib,Cao:2019ygh}. Owing to the unprecedented colliding energy and the fast accumulation of the integrated luminosity, the Large Hadron Collider (LHC) is able to measure the $\tttt$ production~\cite{Sirunyan:2017roi,Sirunyan:2019wxt,Aaboud:2018jsj}. For example, the $\tttt$ signal is observed at the $2.6\sigma$ confidence level at the 13~TeV LHC with an integrated luminosity of $137~{\rm fb}^{-1}$, and the CMS collaboration~\cite{Sirunyan:2019wxt} has published new results on the $\tttt$ cross section,
\begin{equation}
\sigma(\tttt)=12.6^{+5.8}_{-5.2}~{\rm fb}, 
\end{equation}
which is consistent with the tree level prediction in the SM, $\sigma(\tttt)_{\rm SM}=9.6~{\rm fb}$. It yields an upper limit of $\sigma(\tttt)$ at the $95\%$ confidence level as $\sigma(\tttt)\leq  22.5~{\rm fb}$.

\begin{figure}
\centering
\includegraphics[scale=0.45]{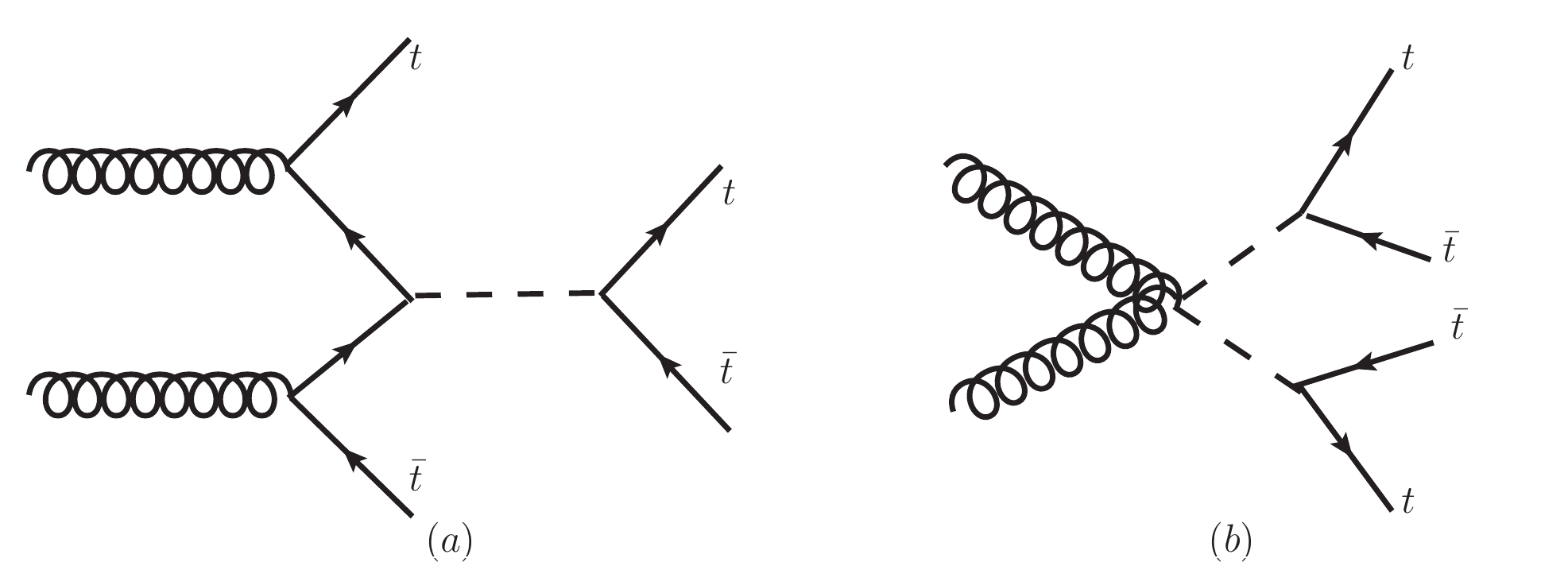}
\caption{Pictorial diagrams of $\tttt$ production in NP models: (a) the associated production of $t\bar{t}X$ with $X\to t\bar{t}$; (b) the pair production of $X\bar{X}$ with $X\to t\bar{t}$. }
\label{FIG:colorneutral}
\end{figure}

In this work we examine the constraint on various NP resonances from the $\tttt$ production. We consider the top-philic NP model in which the NP resonance ($X$) couples only to the top quark. The NP contribution to the $\tttt$ production can be either through the $X\bar{X}$ pair production or through the $X$ production in association with a top-quark pair ($t\bar{t}X$) with a subsequent decay of $X\to t\bar{t}$ or $X\to tt$; see Fig.~\ref{FIG:colorneutral}. Denote $M$ and $\Gamma$ as the mass and width of the $X$ particle, respectively, and $\kappa_X$ as the coupling strength of $X$ to the top quark in the top-philic models. The cross section of the $\tttt$ production can be parametrized as following
\begin{align} 
\label{EQ:highmass}
\sigma^{\rm total}= \sigma^{\rm SM }_{\tttt} + \kappa_X^2 \sigma^{\rm Int}_{\tttt}(M,\Gamma) + \kappa_X^4\sigma^{\rm Res}_{\tttt}(M,\Gamma),
\end{align}
where $\sigma_{\tttt}^{\rm SM}$ ($\sigma_{\tttt}^{\rm Int}$, $\sigma_{\tttt}^{\rm Res}$) denotes the cross section of the SM contribution, the interference between the SM and NP, and the NP contribution alone, respectively. 

The width dependence in Eq.~\ref{EQ:highmass} originates from the propagator of $X$ particle in the intermediate state. One can simplify the cross section when the $X$ particles are on mass shell by using narrow width approximation (NWA)~\cite{HPILUHNL:NWA} which enables us to factorize the $\tttt$ production induced by NP into the $X$ production process and the $X$ decay process unambiguously. The interference between the SM and NP contributions is negligible in the vicinity of $M$, therefore, one can parameterize $\sigma(\tttt)$ as follows:
\bea
&& X\bar{X}~:~\sigma(\tttt)= \sigma(X\bar{X})\times {\rm Br}^2(X\to t\bar{t}),\nn\\
&& t\bar{t}X~:~\sigma(\tttt)=\sigma(t\bar{t}X)\times {\rm Br}(X\to t\bar{t}).
\eea
Note that the resonance $X$ can also decay into a pair of gluons or photons through a top-quark triangle loop in the top-philic model. The loop-induced decays are suppressed by a loop factor $\alpha_{s}/4\pi$ or $\alpha_{e}/4\pi$, resulting in a small branching ratios $\lesssim 1\%$ for a heavy $X$. We assume ${\rm Br}(X\to t\bar{t})=1$ in the study for simplicity.  Therefore, $\sigma(\tttt)$ of the NP process of $pp\to t\bar{t}X\to\tttt$ can be simplified as following:
\begin{align}
\label{EQ:highmass:NAW:l}
\sigma^{\rm NP}_{t\bar{t}X}= \kappa_X^2 \sigma^{\rm Res}_{t\bar{t}X}(M)\times {\rm Br}(X\to t\bar{t}),
\end{align}
while for the process of $ pp\to X\bar{X} \to\tttt$, 
\begin{align} 
\label{EQ:highmass:NAW:r}
\sigma^{\rm NP}_{X\bar{X}}=  \sigma^{\rm Res}_{X\bar{X}}(M)\times {\rm Br}^2(X\to t\bar{t}).
\end{align}
The absence of $\kappa_X$ dependence in $\sigma^{\rm NP}_{X\bar{X}}$ is due to the fact that only gauge interaction involves in the $X\bar{X}$ pair production.

As the width $\Gamma$ increases dramatically with $\kappa_X$, the width effect can be sizable and then the NWA might not be valid. Equations~\ref{EQ:highmass:NAW:l} and~\ref{EQ:highmass:NAW:r} are meant only to illustrate the dependence of $\sigma(\tttt)$ on NP parameters, and we keep the width effect of $X$ particle thoroughly in our calculation. We adopt the CERN LEP line-shape prescription of a resonance state~\cite{Cao:2004yy} and write
the $X$ propagator as
\beq
\frac{1}{(p^{2}-M^{2})+iM\Gamma \left(p^2/{M^2}\right)}
\eeq
which makes a distortion on the dispersion relation at $\Gamma^2/2M^2$ level. A comparison between the full width effect and NWA is present later.

When the $X$ particle is below the top quark pair mass threshold ($M\leq 340~{\rm GeV}$)  or too heavy to be produced directly at the LHC, the $X$ particle in the intermediate state can never be on mass shell. Therefore, the $\tttt$ production then depends on $\kappa_X$ and $M$, 
\begin{align} 
\label{EQ:lowmass}
\sigma^{\rm total}= \sigma^{\rm SM }_{\tttt} + \kappa_X^2 \sigma^{\rm Int}_{\tttt}(M) + \kappa_X^4\sigma^{\rm Res}_{\tttt}(M).
\end{align}
For colored resonances the process in Fig.~\ref{FIG:colorneutral}(a) is dominant for a light $X$ as it consists of more Feynman diagrams than the $X\bar{X}$ pair production (b). On the other hand, the process in Fig.~\ref{FIG:colorneutral}(b) dominates when $m_X>2m_t$.

~\\
\noindent\textbf{2. New Physics Resonances. }

Now consider the LHC search on NP resonances. Rather than focusing on specific NP models, we consider various simplified NP models which extend the SM with an additional NP resonance. The effective Lagrangians are listed as follows:
\begin{itemize}
\item A color singlet scalar ($\mathcal{S}$), e.g.,~\cite{Dev:2014yca,Alvarez:2016nrz,Alvarez:2019uxp},
\begin{align}
\mathcal{L}\supset -\bar{t}(a_t+i\gamma_5 b_t)t\mathcal{S}\,, 
\end{align}
where $a_t=0$ and $b_t=0$ corresponds to a CP-odd scalar ($A$) and a CP-even scalar ($H$), respectively, and the case of $a_t\neq 0$ and $b_t\neq 0$ represents a CP-mixture scalar; 

\item a color octet scalar ($\mathcal{S}_8$), e.g., sgluon~\cite{Calvet:2012rk,Darme:2018dvz},
\begin{align}
\mathcal{L}\supset \text{Tr} [(D_\mu \mathcal{S}_8)^\dagger(D^\mu \mathcal{S}_8)]-\mathcal{S}_8^A\overline{t}(g_8^a+i\gamma_5 g_8^b)T^At,
\end{align}
where $T^A$ stands for the $SU(3)_C$ generator;

\item a color octet vector ($\mathcal{V}_{8\mu}$), e.g., axigluon~\cite{Antunano:2007da,Chivukula:2010fk}, color octet boson~\cite{Ferrario:2008wm,AguilarSaavedra:2011ck}, KK gluon~\cite{Guchait:2007jd},
\begin{align}
\mathcal{L}\supset& -\frac{1}{2}D_\nu \mathcal{V}_{8\mu}^A(D^\nu \mathcal{V}_{8}^{\mu A}-D^\mu \mathcal{V}_{8}^{\nu A})\nonumber\\
&-\frac{1}{2}g_sf^{ABC}G_{\mu\nu A}\mathcal{V}_{8\mu}^B\mathcal{V}_{8\mu}^C\nonumber\\
&+\mathcal{V}_{8\mu}^A\overline{t}(g_8^V\gamma^\mu+g_8^A\gamma^\mu\gamma_5)T^At\,, 
\end{align}
where $g_8^V$ and $g_8^A$ denotes the vector and axial vector coupling, respectively; 

\item a color singlet vector ($\mathcal{V}_\mu$)\cite{Greiner:2014qna,Kim:2016plm},
\begin{align}
\mathcal{L}\supset V_{\mu}\overline{t}(g^V\gamma^\mu +g^A\gamma^\mu \gamma^5)t\,,
\end{align}
where $g^V$ and $g^A$ presents the vector and axial vector coupling,  respectively; 

\item a color sextet and EW singlet scalar $\mathcal{S}_6$~\footnote{The color sextet scalar could be an EW singlet or triplet under $SU(2)_L$, which couples to right-handed or left-handed fermions in the SM, respectively. As an EW triplet and color sextet, the scalar could induce rich collider signatures such as $b\bar{b}\to t\bar{t}$. For simplicity we focus on the case of $S_6$ being a EW singlet here. }, e.g.,~color sextet scalar~\cite{Chen:2008hh,Berger:2010fy,Zhang:2010kr},
\begin{align}
\mathcal{L}&\supset~ \text{Tr} [ (D_\mu \mathcal{S}_6)^\dagger(D^\mu \mathcal{S}_6)] \nonumber\\
&+~g_6 \mathcal{S}_6^A\overline{t}_R\overline{K}^At^C_R+\mbox{h.c.}\, ,
\end{align}
where $K^A$ stands for the Clebsh-Gordon coefficient.
\end{itemize}
Equipped with the effective Lagrangians shown above, we are ready to check the validation of the NWA. The decay width of NP resonances into a pair of top quarks, $\Gamma(X)\equiv \Gamma(X\to t\bar{t})$ or $\Gamma(X\to tt)$, are 
\begin{align}
\Gamma(\mathcal{S}) =& \frac{3M}{8\pi} \left[(a_t)^2 \beta_t^2+(b_t )^2 \right ] \beta_t, \label{EQ:width:S}\\
\Gamma(\mathcal{S}_8) =& \frac{3M}{48\pi} \left[(g_8^a)^2 \beta_t^2+(g_8^b)^2 \right ] \beta_t,\label{EQ:width:S_8} \\
\Gamma(\mathcal{V}) =& \frac{M}{4\pi}\left [ (g^{V})^2 (1+ 2 \frac{m_t^2}{M^2}) +(g^{A})^2 \beta_t^2  \right ] \beta_t , \label{EQ:width:V} \\
\Gamma(\mathcal{V}_{8} ) =& \frac{M}{24\pi} \left[ (g_{8}^{V})^2  (1 +2  \frac{m_t^2}{M^2} )+(g_{8}^{A})^2 \beta_t^2   \right ] \beta_t,\label{EQ:width:V_8}\\
\Gamma(\mathcal{S}_6^A) =& \frac{g_6^2 M}{8\pi} \left(1-2\frac{m_t^2}{M^2}\right)\beta_t, \label{EQ:width:S_6}
\end{align}
where $\beta_t=\sqrt{1-4m_t^2/M^2}$ is the velocity of top quark.
As the colored resonance exhibits a narrower width in comparison with the color neutral objects, 
e.g. the width of a color-octet (sextet) resonance is $1/6$ ($1/3$) of a color neutral resonance, respectively, we examine the width-to-mass ratio of color-neutral resonances below.

\begin{figure}
\centering
\includegraphics[scale=0.6]{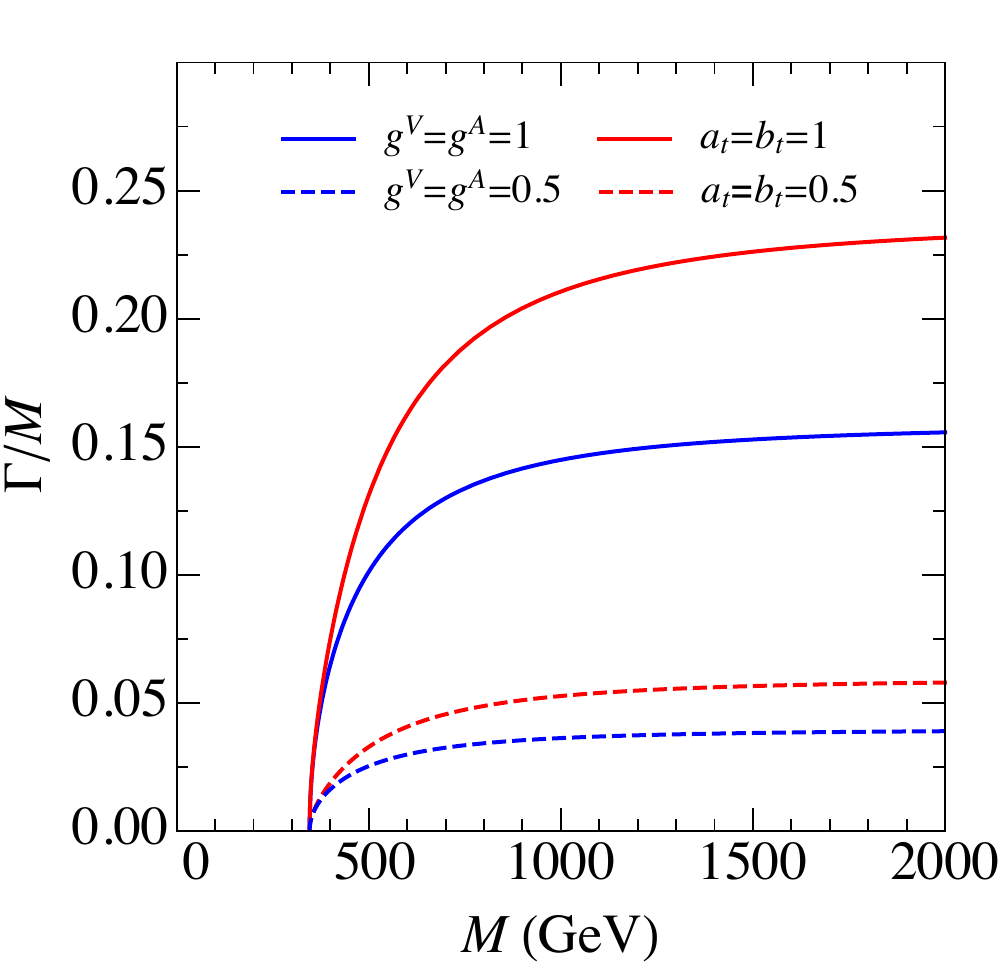}
\caption{Ratio $\Gamma/M$ as a function of $M$ for a color-neutral scalar (red) and color-neutral vector (blue). }
\label{FIG:width}
\end{figure}

Figure~\ref{FIG:width} plots the ratio $\Gamma/M$ as a function of $M$ for a color singlet scalar (red) and color singlet vector (blue). For demonstration we choose two sets of coupling parameters.
When the strength of its coupling to top quark is large,  a heavy color-singlet scalar exhibits a large width, e.g. $\Gamma/M\simeq 23\%$  in the region of $M\sim 2000~{\rm GeV}$; see the red-solid curve. However, for such a heavy scalar, the $\sigma(\tttt)$ is highly suppressed such that the large width leads to a mild effect in the constraints on NP resonance. The width effect turns to be sizable in the $\tttt$ production in the region of $M\sim 800~{\rm GeV}$. On the other hand, the width-to-mass ratio of the colored scalar and vector do not exceed $5\%$ such that the NWA works well and the interference effect can be safely dropped.
In this study we include the full width effect in the calculation to constrain the NP resonance at the LHC and a comparison of the full width and NWA is made in the end of the section.  

\begin{figure}[b]
\includegraphics[scale=0.7]{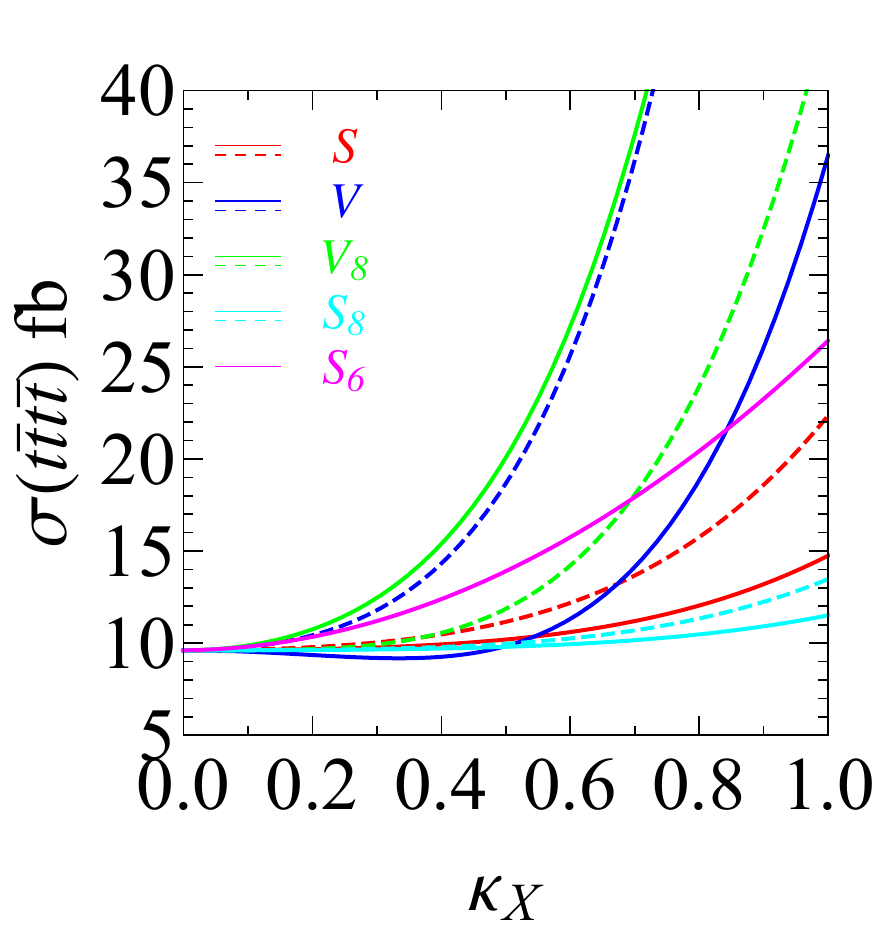}
\caption{The $\tttt$ production cross section as a function of $\kappa_X$ in the extended simplified models for $M=300~{\rm GeV}$. The solid lines represent the scalar or vector couplings while the dashed curves the pseudo-scalar or axial-vector couplings.}
\label{FIG:300}
\end{figure}

Next we show the comparison of the four top quark production rate in those simplified models. First we focus on a light resonance, namely its mass is smaller than $2m_t$ and we fix the mass of 300 GeV. For such a light resonance, it can only contribute to four top quark production via off-shell effects and its width effect can be neglected in the cross section calculation. It means we can parametrize the cross section in terms of its couplings.  We assume that both the SM and NP processes exhibit the same $K$-factor of 1.58~\cite{Frederix:2017wme}.
The cross section of $\tttt$ production induced by 300~GeV NP resonances at the 13~TeV LHC are given by 
 \begin{align}
\sigma^{\rm total}_\mathcal{S}= &9.608 + 1.414 (a_t)^2 + 3.999 (b_t)^2 + \nonumber\\
&  3.724 (a_t)^4  + 10.771 (a_t b_t)^2 + 8.750 (b_t)^4, \nonumber\\
\sigma^{\rm total}_\mathcal{V}= &9.608 - 7.728 (g^V)^2 + 17.394 (g^A)^2 + \nonumber\\
 & 34.604 (g^V)^4  + 32.648 (g^V g^A)^2 + 75.216 (g^A)^4, \nonumber \\
 \sigma^{\rm total}_{\mathcal{S}_8}=& 9.608 +  0.378(g_8^{a})^2 + 0.623(g_8^{b})^2+\nonumber\\
&   1.522(g_8^{a})^4  +  6.017(g_8^{a}g_8^{b})^2 + 3.244(g_8^{b})^4, \nonumber\\
\sigma^{\rm total}_{\mathcal{V}_8}=& 9.608 +  25.895(g_8^{V})^2 +  0.412(g_8^{A})^2+ \nonumber\\
&  63.398(g_8^{V})^4  +  82.039(g_8^{V} g_8^{A})^2 +  34.292(g_8^{A})^4, \nonumber \\
\sigma^{\rm total}_{\mathcal{S}_6}=& 9.608 +  0.373(g_6)^2 + 16.435(g_6)^4. \nonumber
 \end{align}
Throughout the paper all the cross sections are in the unit of femtobarn (fb). The magnitude of the coefficients before effective coupling combinations reveals the relative size of interference and NP contribution in comparison with the SM prediction (i.e. the constant term in above equation). 
We also plot the $\sigma(\tttt)$ as a function of effective couplings (denoted by $\kappa_X$) in Fig.~\ref{FIG:300}, and for simplicity we consider one coupling at a time. The solid lines represent the scalar or vector couplings while the dashed curves denote the pseudo-scalar or axial-vector couplings. Obviously, when the coupling approaches to 1, the NP contribution alone, i.e. the terms proportional to $\kappa_X^4$, tends to dominate the production cross section. For a medium $\kappa_X$, the axial-vector and pseduo-scalar couplings enhance the cross section sizably owing to the interference with the SM contribution. 

Second, we consider the production of an 800~GeV resonance in the $\tttt$ production. For a color neutral resonance, the $\tttt$ production is dominated by the $t\bar{t}X$ association production with $X\to t\bar{t}$; while for a colored resonance, the $X\bar{X}$ pair production through pure QCD with $X\to t\bar{t}$ in consequence overwhelmingly dominates the $\tttt$ production assuming ${\rm Br}(X\to t\bar{t})=1$. Under the NWA approximation, the $\tttt$ production cross sections read as  
\begin{align}
\sigma^{\rm total}_\mathcal{S}=& 9.608 + 5.926 (a_t)^2 + 7.076 (b_t)^2, \\
\sigma^{\rm total}_\mathcal{V}= & 9.608 + 35.804 (g^V)^2 - 1.830 g^Vg^A + \nn \\
&35.804 (g^A)^2, \label{eq:heavy} \\
\sigma^{\rm total}_{\mathcal{S}_8}=& 189.477 +  20.166 (g_8^{a})^2 + 19.4288 (g_8^{b})^2, \\
\sigma^{\rm total}_{\mathcal{V}_8}=& 6792.253 +  196.729 (g_8^{V})^2 +  189.482 (g_8^{A})^2, \\
\sigma^{\rm total}_{\mathcal{S}_6}=& 349.305 + 63.197 (g_6)^2. 
\end{align}
The cross section depends only on the quadratic power of the effective coupling $\kappa_X$ 
as the decay branching ratio of $X\to t\bar{t}$ is assumed to be 1 and independent of $\kappa_X$.
The constant terms represent the sum of the SM contribution and the $X\bar{X}$ production, if applicable.

Now we are ready to explore constraints on various top-philic resonances from the recent result of the $\tttt$ production reported by the CMS collaboration, $\sigma(\tttt)=12.6^{+5.8}_{-5.2}~{\rm fb}$~\cite{Sirunyan:2019wxt}. As the sensitivity of the $\tttt$ production to $\kappa_X$ highly depends on the production channel and the mass of the resonance $M$, we consider both light resonances and heavy resonances in this study.

\begin{figure}
\centering
\includegraphics[scale=0.28]{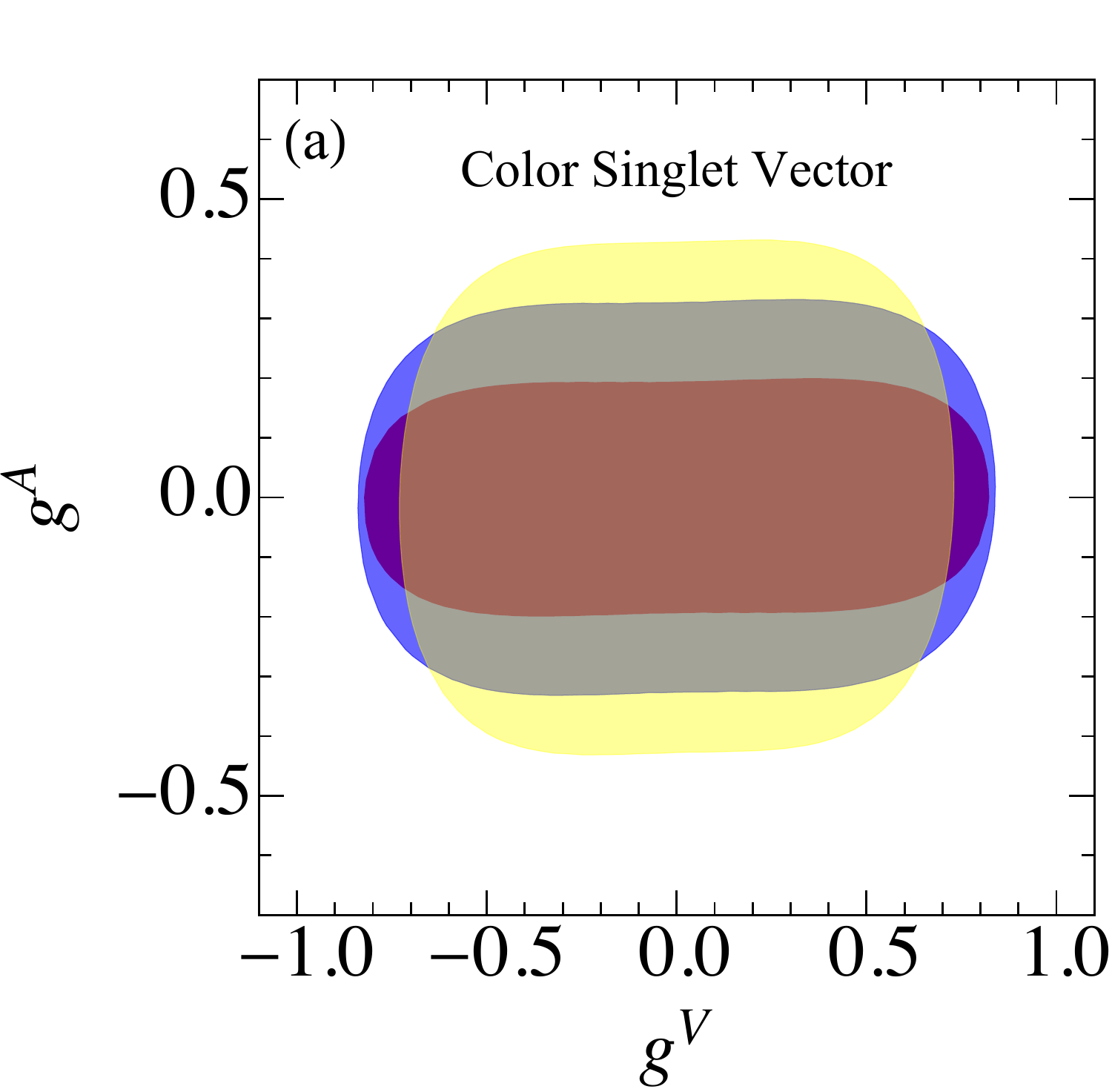}~~~\includegraphics[scale=0.29]{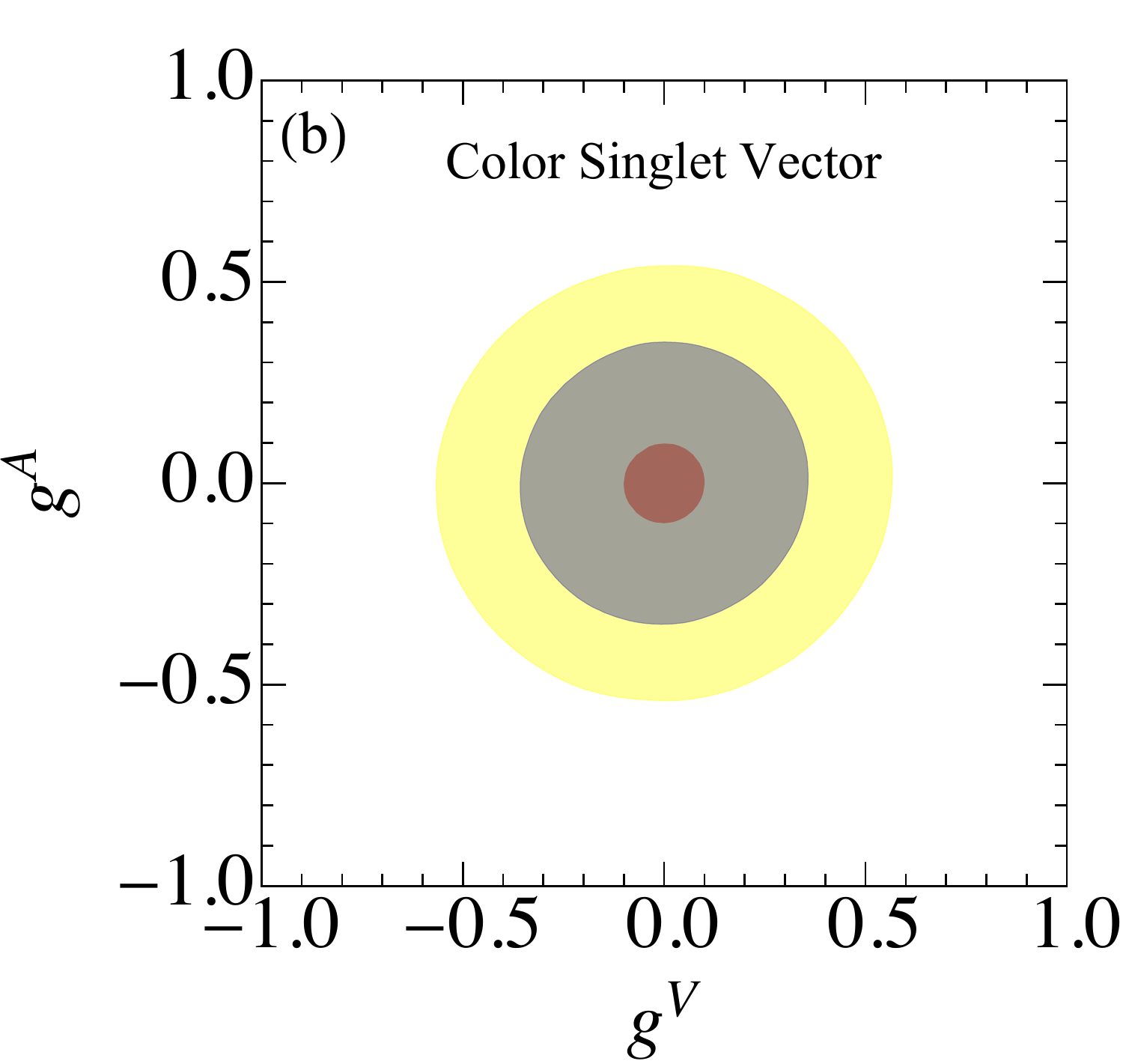}
\includegraphics[scale=0.28]{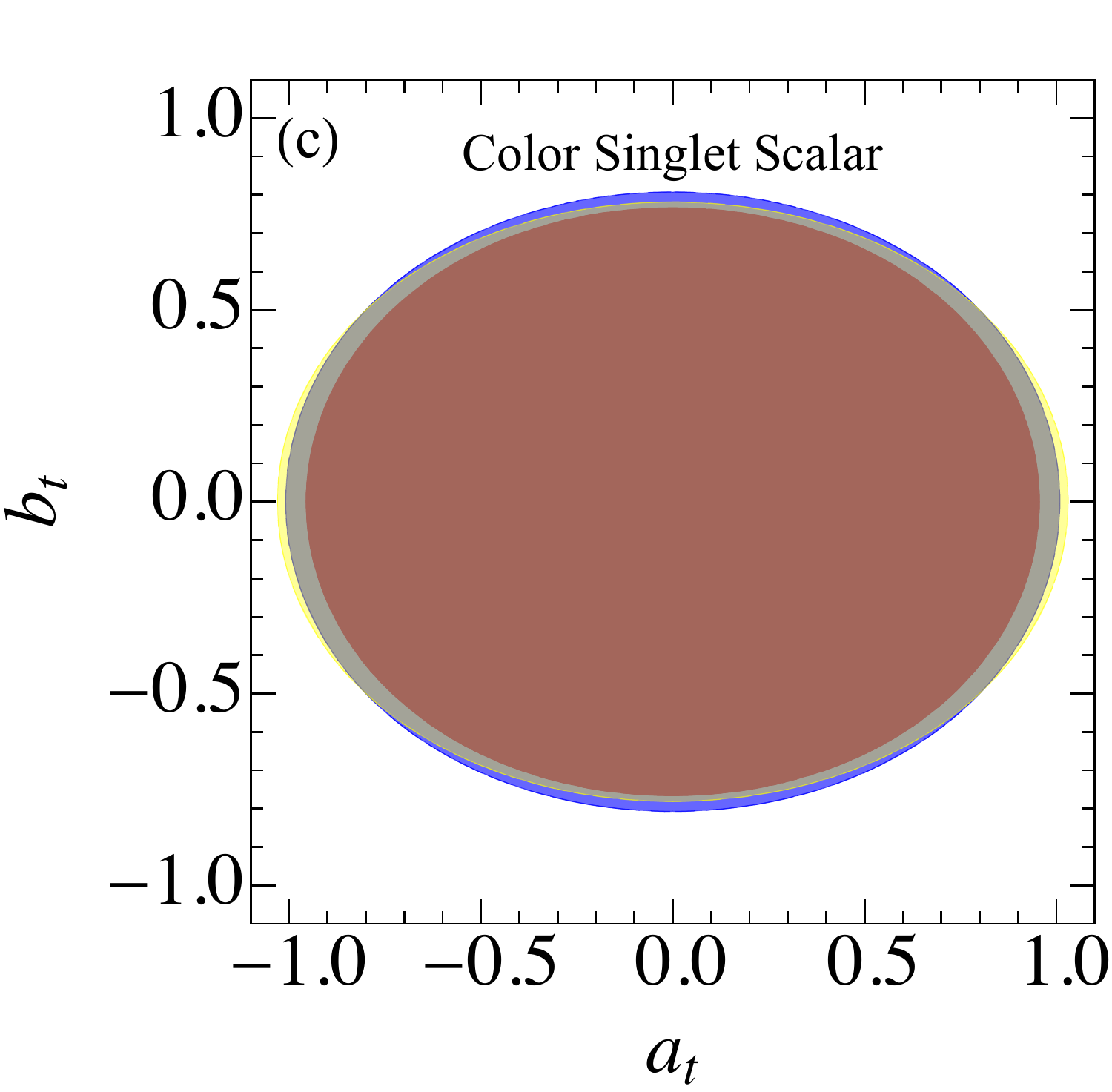}~~~\includegraphics[scale=0.28]{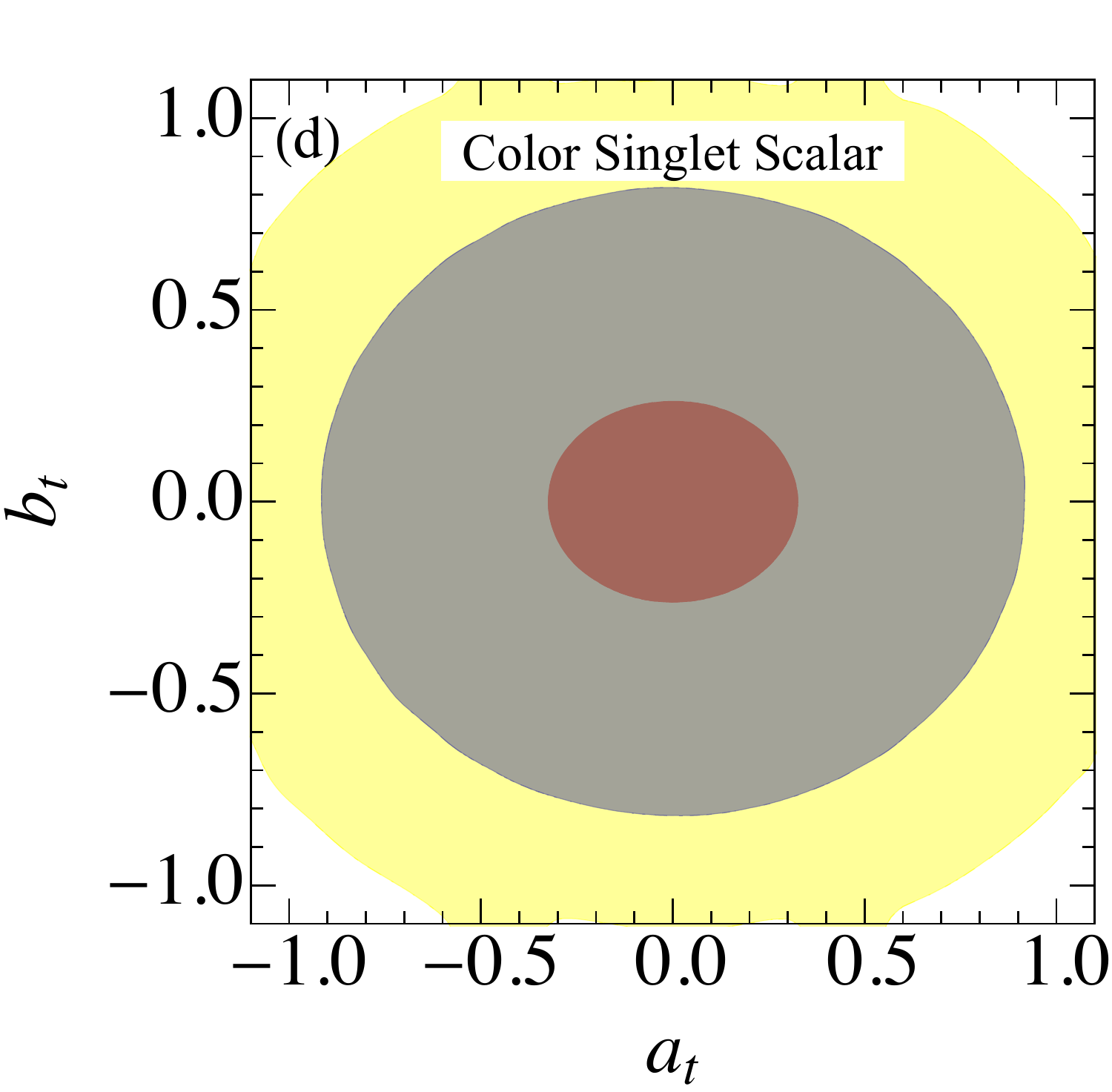}
\caption{Allowed region for a color-neutral vector $\mathcal{V}$ (a, b) and scalar $\mathcal{S}$ (c, d). The red (blue, yellow) region in (a, c) presents the resonance with a mass of 100 (200, 300) GeV, respectively, and in (b, d) with a mass of 350 (800, 1000) GeV, respectively.}
\label{Fig:vectorresonance}
\end{figure}

We begin with the case of a color singlet vector boson $\mathcal{V}$. Fig.~\ref{Fig:vectorresonance}(a) displays the allowed parameter region in the plane of ($g^A$, $g^V$) for a light color-neutral vector. For illustration we consider three benchmark masses, 100~GeV (red), 200~GeV (blue) and 300~GeV (yellow). The axial-vector coupling is constrained more than the vector coupling as the axial-vector contribution is enhanced by $m_t^2/m_{\mathcal{V}}^2$ for a light resonance; for example,  $|g^A| \lesssim 0.2$ and $|g^V|\lesssim 1$. Figure~\ref{Fig:vectorresonance}(b) shows the allowed parameter space of a heavy resonance for three benchmark masses, 350~GeV (red), 800~GeV (blue) and 1000~GeV (yellow). The major contribution to the $\tttt$ production is from the $t\bar{t}\mathcal{V}$ production whose cross section depends on the quadratic power of $g^{V/A}$'s; see $\sigma^{\rm total}_{\mathcal{V}}$ for $M=800~{\rm GeV}$ in Eq.~\ref{eq:heavy}.  It leads to a circle parameter space centering around $g^A=g^V=0$. The $\sigma(\tttt)$ decreases dramatically with $M$ so as to weaken the bound; for example, $|g^{V,A}|\lesssim 0.5$ for $M=1000~{\rm GeV}$ while $|g^{V,A}|\lesssim 0.2$ for $M=350~{\rm GeV}$. 

Figure~\ref{Fig:vectorresonance}(c) and (d) shows the allowed parameter space for a light and heavy color-neutral scalar $\mathcal{S}$, respectively. First, we observe a similar pattern as the color-neutral vector but with weaker bounds. Note that the contribution of the scalar interaction ($a_t\neq 0$, $b_t=0$) is suppressed by the top-quark velocity, therefore, the scalar interaction is less constrained. Second, owing to the small production rate, the TeV scalar is loosely bounded, e.g. $\sqrt{a^2_t+b^2_t}\sim 1.5$; see the yellow region in Fig.~\ref{Fig:vectorresonance}(d). The NP scalar exhibits a large width for such a large coupling, e.g. $\Gamma/M\sim 30\%$ and cannot be treated as a fundamental particle.

\begin{figure}[b]
\centering
\includegraphics[scale=0.285]{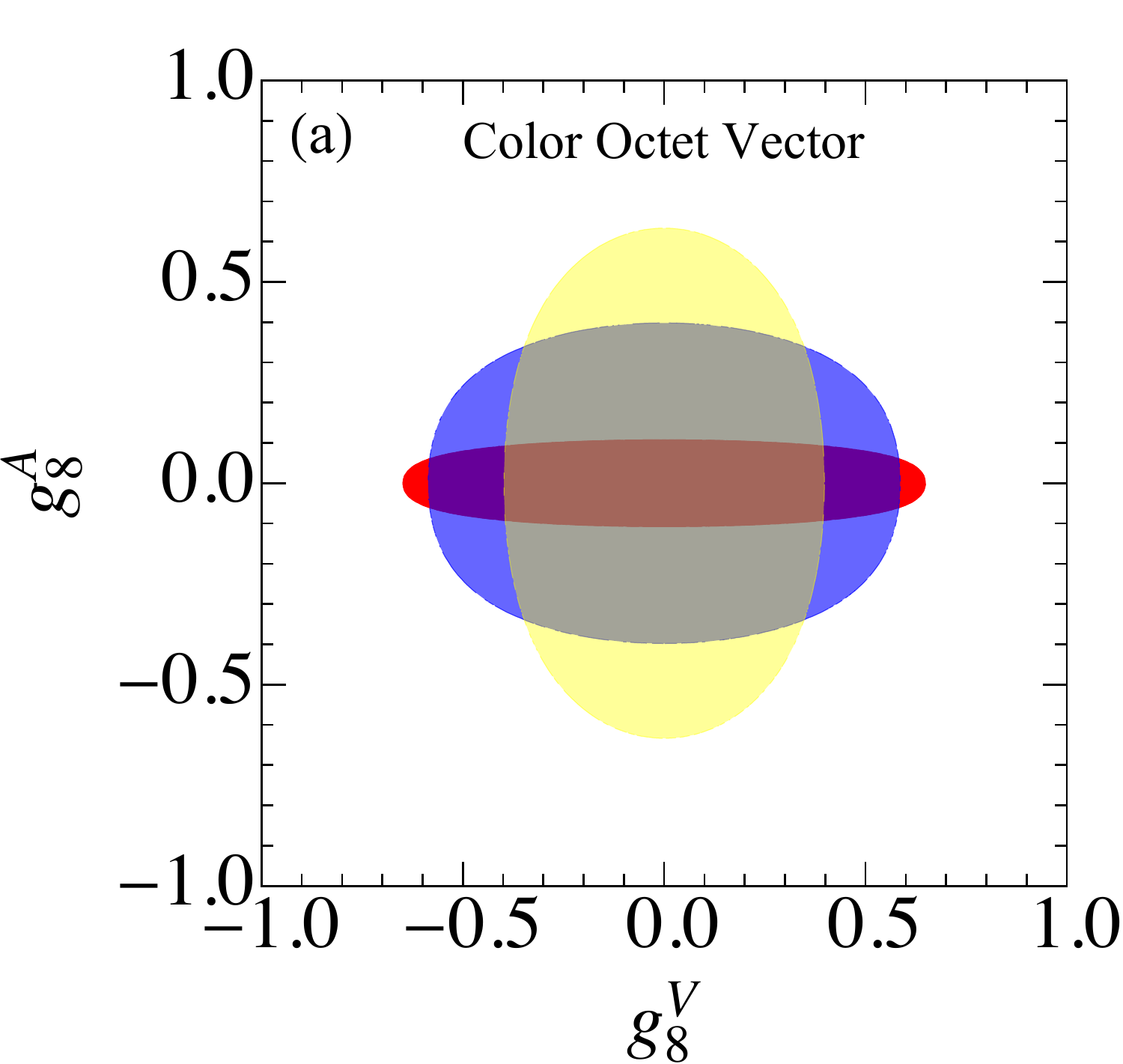}
\includegraphics[scale=0.27]{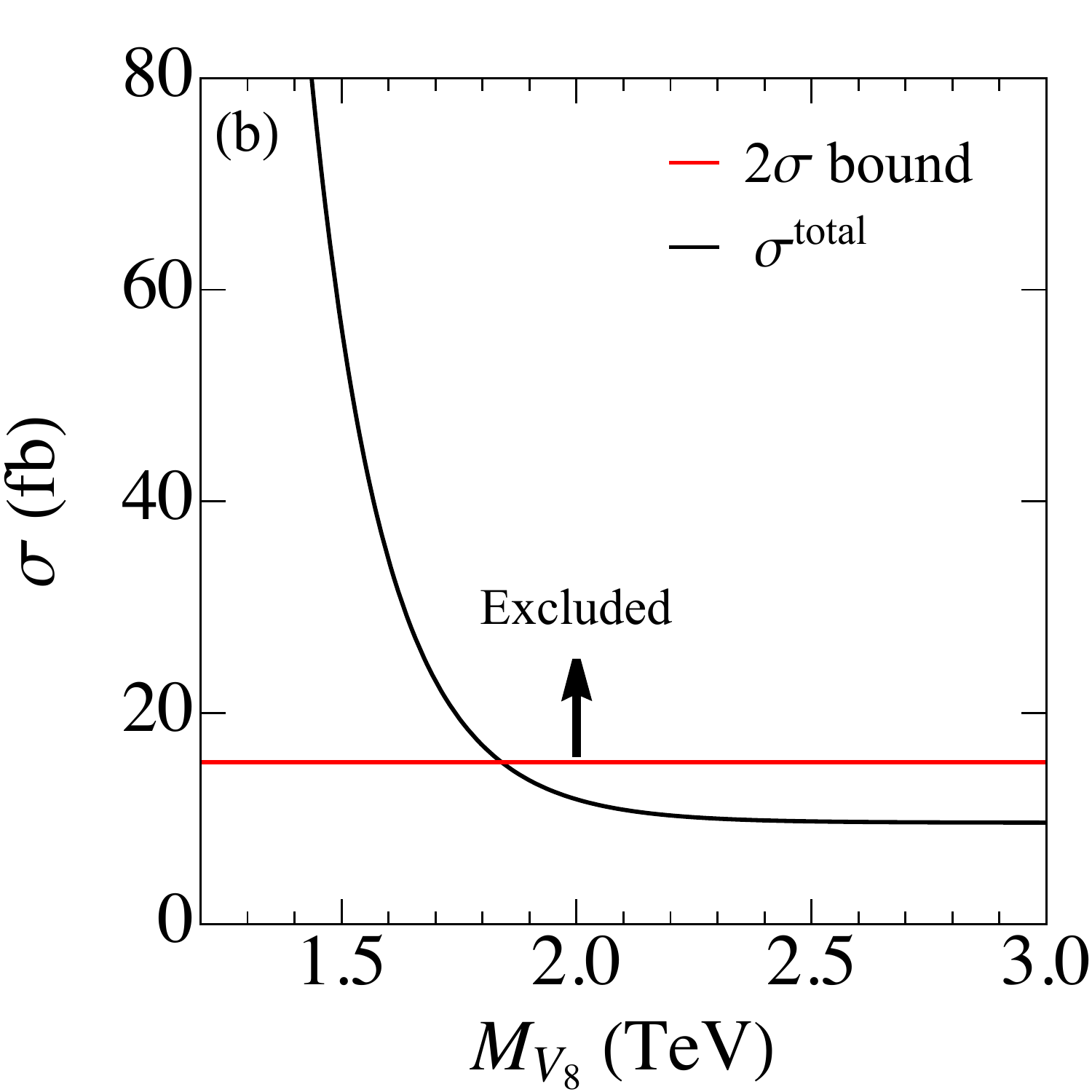}
\caption{Allowed region for a light (a) and heavy (b) color-octet vector boson $\mathcal{V}_8$. The red (blue, yellow) region in (a) denotes a vector boson with mass of 100 (200, 300) GeV, respectively. The black curve in (b) denotes $\sigma(\tttt)$ as a function of $m_{\mathcal{V}_8}$ while the red curve represents the exclusion limit on $\sigma(\tttt)$ at the $2\sigma$ confidence level.}
\label{Fig:v8resonance}
\end{figure}

Next, we consider a color octet vector boson $\mathcal{V}_8$.  
Figure~\ref{Fig:v8resonance}(a) plots the allowed parameter space in the plane of ($g_8^V$, $g_8^A$) for three benchmark masses of a light $\mathcal{V}_8$. The red (blue, yellow) region in (a) denotes a vector boson with mass of 100 (200, 300) GeV, respectively. The contribution of the axial-vector current interaction in the four-top production is enhanced by $m_t^2/m_{\mathcal{V}_8}^2$, therefore, the lighter the $\mathcal{V}_8$ is, the larger the $\sigma(\tttt)$ can be. For example, the bound of a 100~GeV vector (red) is much tighter than the bound of a 300~GeV vector (yellow). However, the enhancement disappears when $m_{\mathcal{V}_8}\sim 300~{\rm GeV}$ and the interference with the SM gluon contribution, i.e. the vector coupling, plays a leading role. As a result, the $g_8^V$ is constrained more tightly than the $g_8^A$ for a 300~GeV resonance. When the vector boson mass exceeds $2m_t$, as the vector boson carries color charge, the four top production is dominated by the $\mathcal{V}_8\mathcal{V}_8$ production with a subsequent decay of $\mathcal{V}_8\to t\bar{t}$. The production rate depends only on the color charge of $\mathcal{V}_8$ but not on the $g_8^{V}$ and $g_8^A$. The black curve in Fig.~\ref{Fig:v8resonance}(b) shows $\sigma(\tttt)$, i.e. $\sigma(\mathcal{V}_8\mathcal{V}_8)$ with ${\rm Br}(\mathcal{V}_8\to t\bar{t})=1$, at the LHC. The red line denotes the exclusion bound at the $2\sigma$ confidence level which shows that the mass of $\mathcal{V}_8$ is larger than 1.82~TeV.

\begin{figure}
\centering
\includegraphics[scale=0.27]{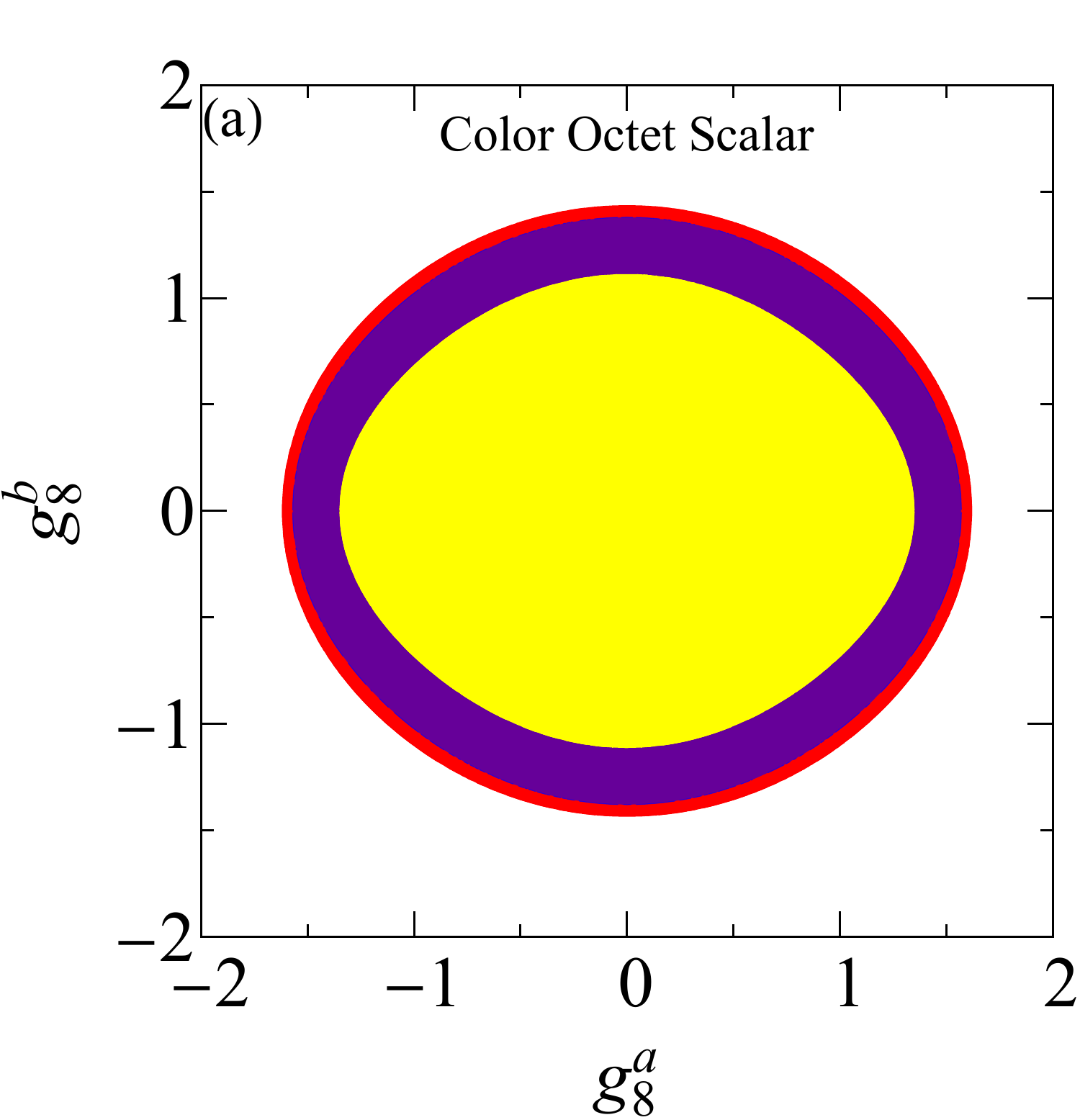}~
\includegraphics[scale=0.274]{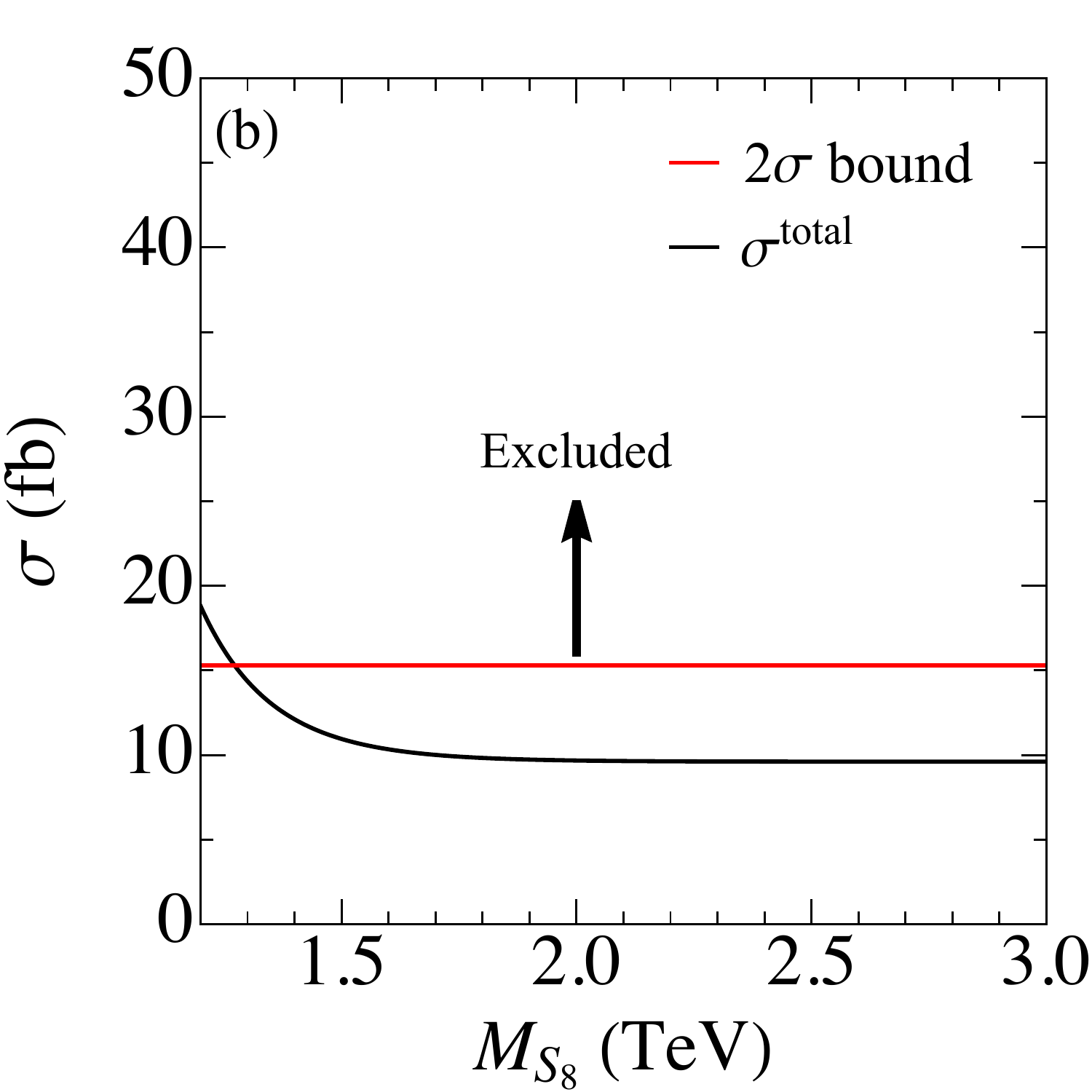}
\caption{Allowed region for a light (a) and heavy (b) color octet scalar $\mathcal{S}_8$. The red (blue, yellow) region represents the mass of 100 (200, 300)~GeV, respectively.  The black curve in (b) denotes $\sigma(\tttt)$ as a function of $M_{\mathcal{S}_8}$ while the red curve represents the exclusion limit on $\sigma(\tttt)$ at the $2\sigma$ confidence level.}
\label{Fig:s8resonance}
\end{figure}

Now we consider a color octet scalar $\mathcal{S}_8$. Figure~\ref{Fig:s8resonance}(a) plots the allowed parameter space of $g_8^a$ and $g_8^b$ where the red (blue, yellow) region represents the $\mathcal{S}_8$ mass of 100 (200, 300)~GeV, respectively. 
When $m_{\mathcal{S}_8}\sim 300~{\rm GeV}$, the QCD pair production becomes important and enlarges the production cross section, yielding a smaller region; see the yellow oval. Once going above the top-quark pair threshold, say $m_{\mathcal{S}_8}>2m_t$, the current data demands $m_{\mathcal{S}_8}>1.19~{\rm TeV}$; see Fig.~\ref{Fig:s8resonance}(b). 

\begin{figure}
\centering
\includegraphics[scale=0.39]{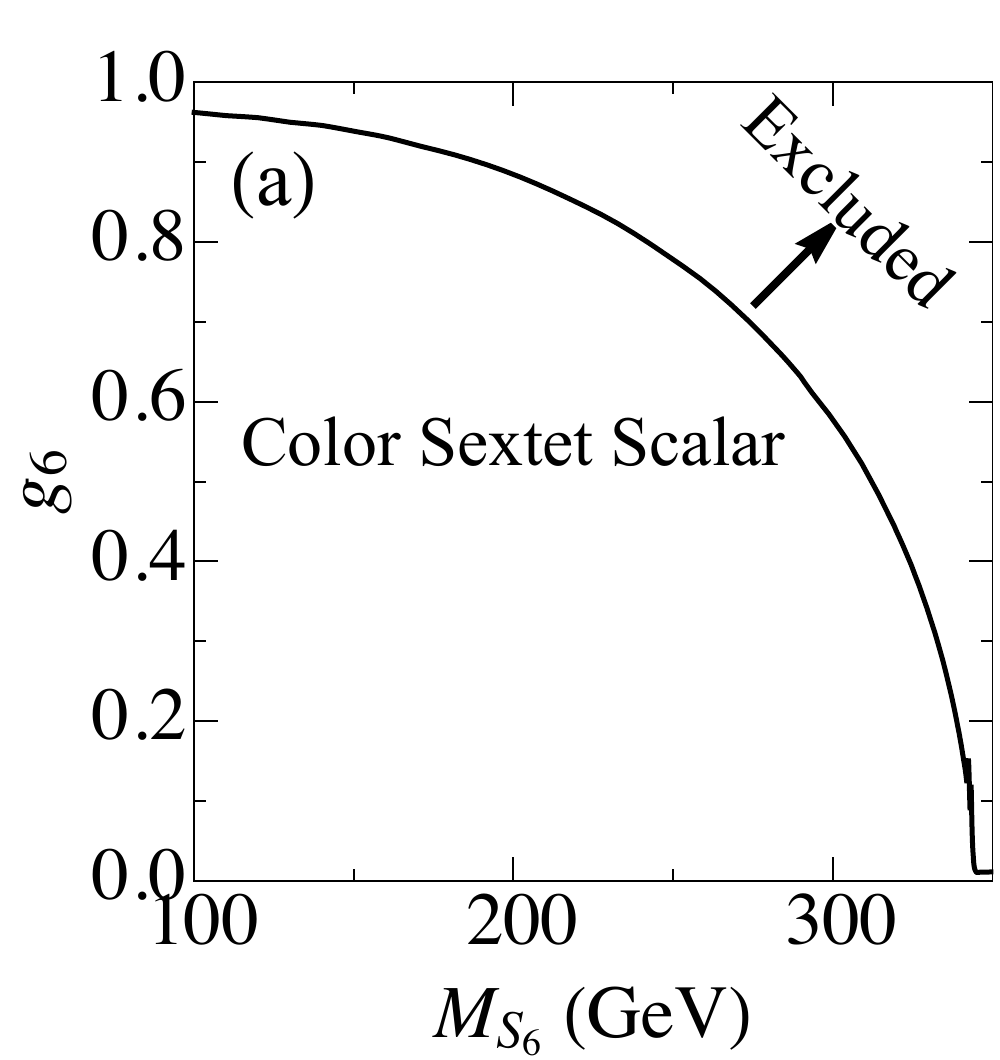}~~~
\includegraphics[scale=0.27]{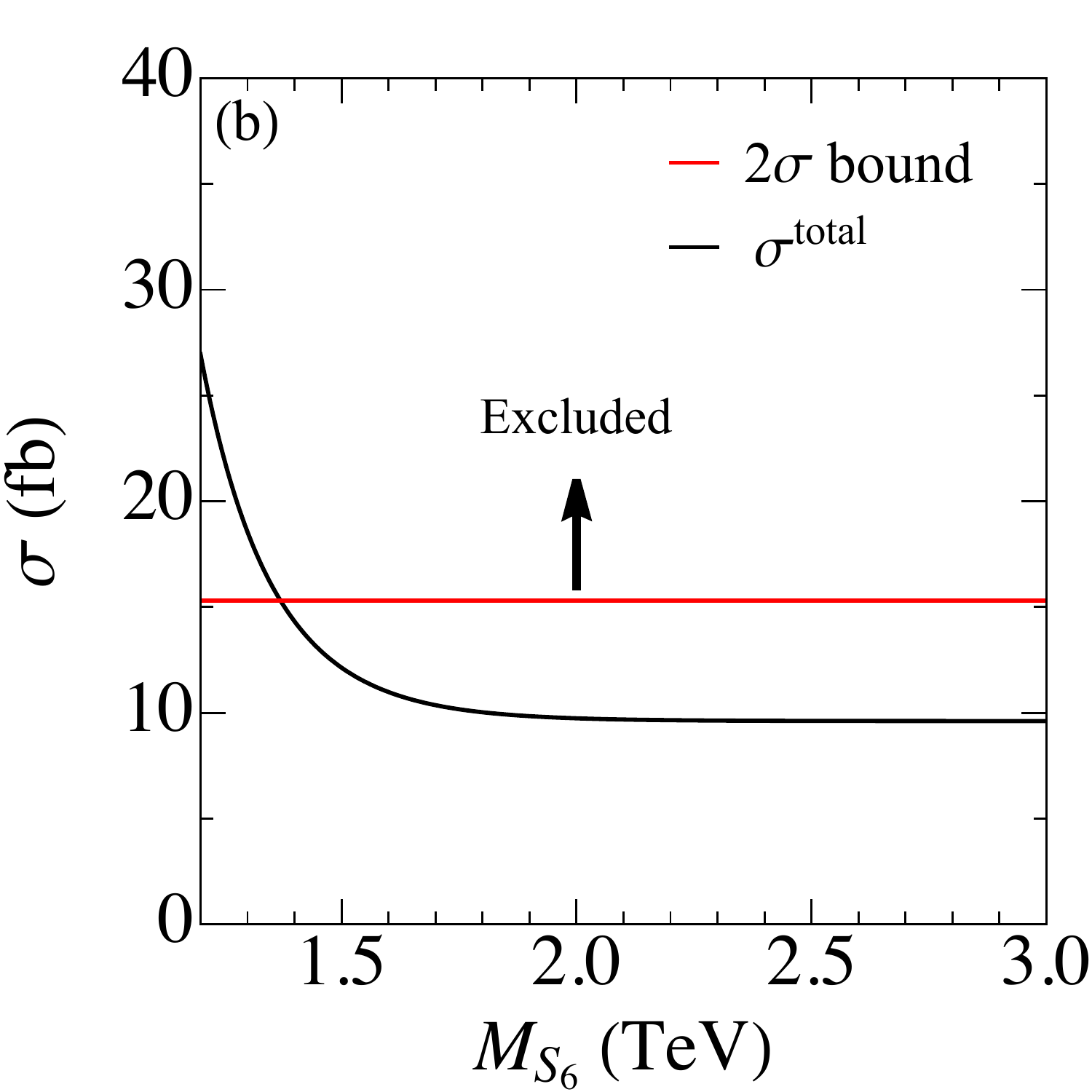}
\caption{\label{Fig:s6resonance}Limits on the color sextet scalar coupling and mass from current four top quark production constraints. }
\end{figure}

Last but not least, we discuss the case of a color sextet scalar $\mathcal{S}_6$~\cite{Chen:2008hh,Berger:2010fy,Zhang:2010kr}. 
The production depends on $m_{\mathcal{S}_6}$ and $g_6$. Again, we consider both light and heavy scalars. Figure~\ref{Fig:s6resonance}(a) displays the bound on a light color-sextet scalar where the region above the black curve is excluded by the current data. Once $m_{\mathcal{S}_6}>2m_t$ the $\mathcal{S}_6 \mathcal{S}^\dagger_6$ pair production dominates the $\tttt$ production such that the $\sigma(\tttt)$ does not depend on $g_6$ at all; see the black curve in Fig.~\ref{Fig:s6resonance}(b). To respect the current bound on $\sigma(\tttt)$ (red), the mass of $\mathcal{S}_6$ has to be larger than 1.38~TeV.

\begin{figure}
\centering
\includegraphics[scale=0.28]{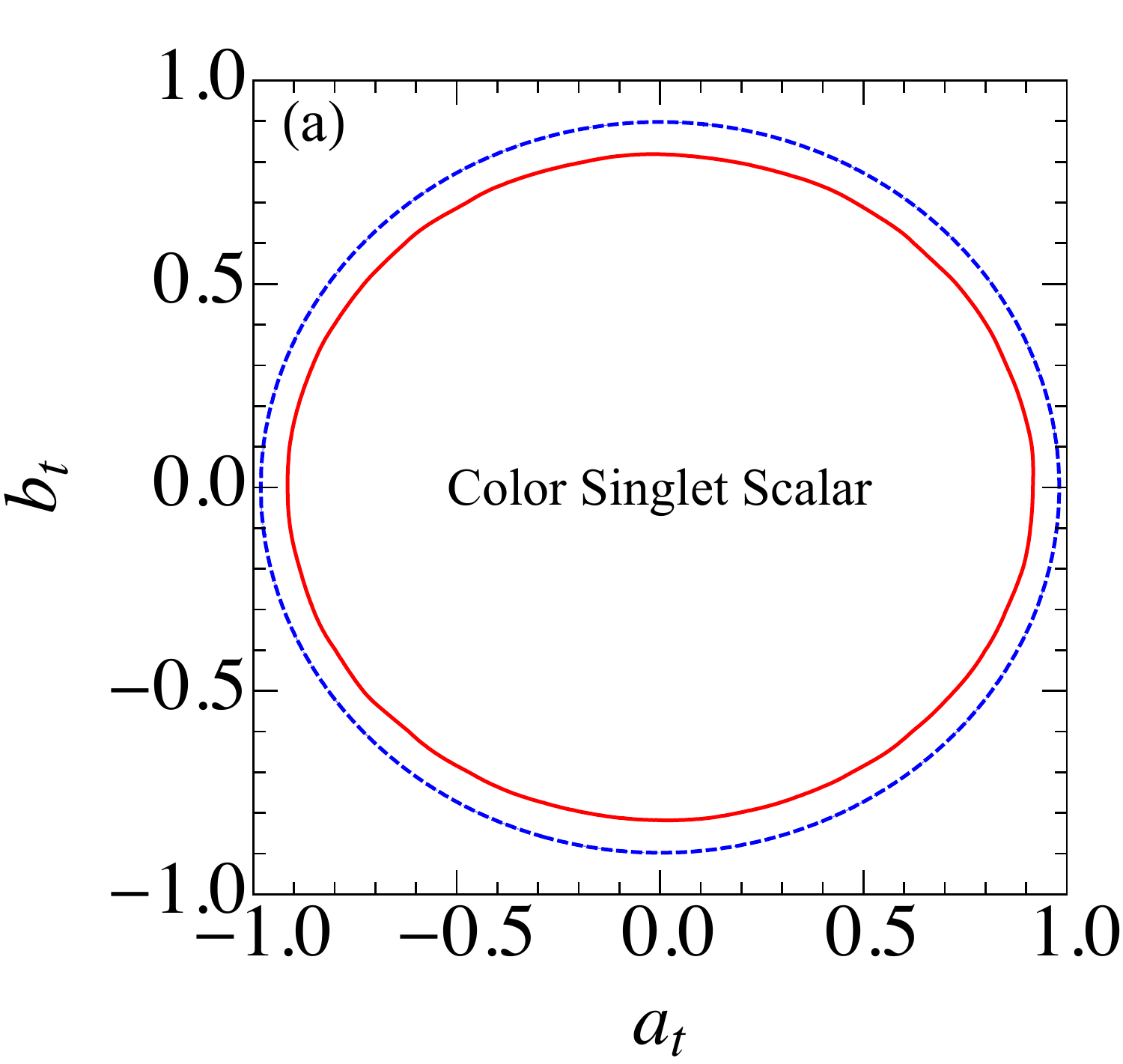}~~
\includegraphics[scale=0.28]{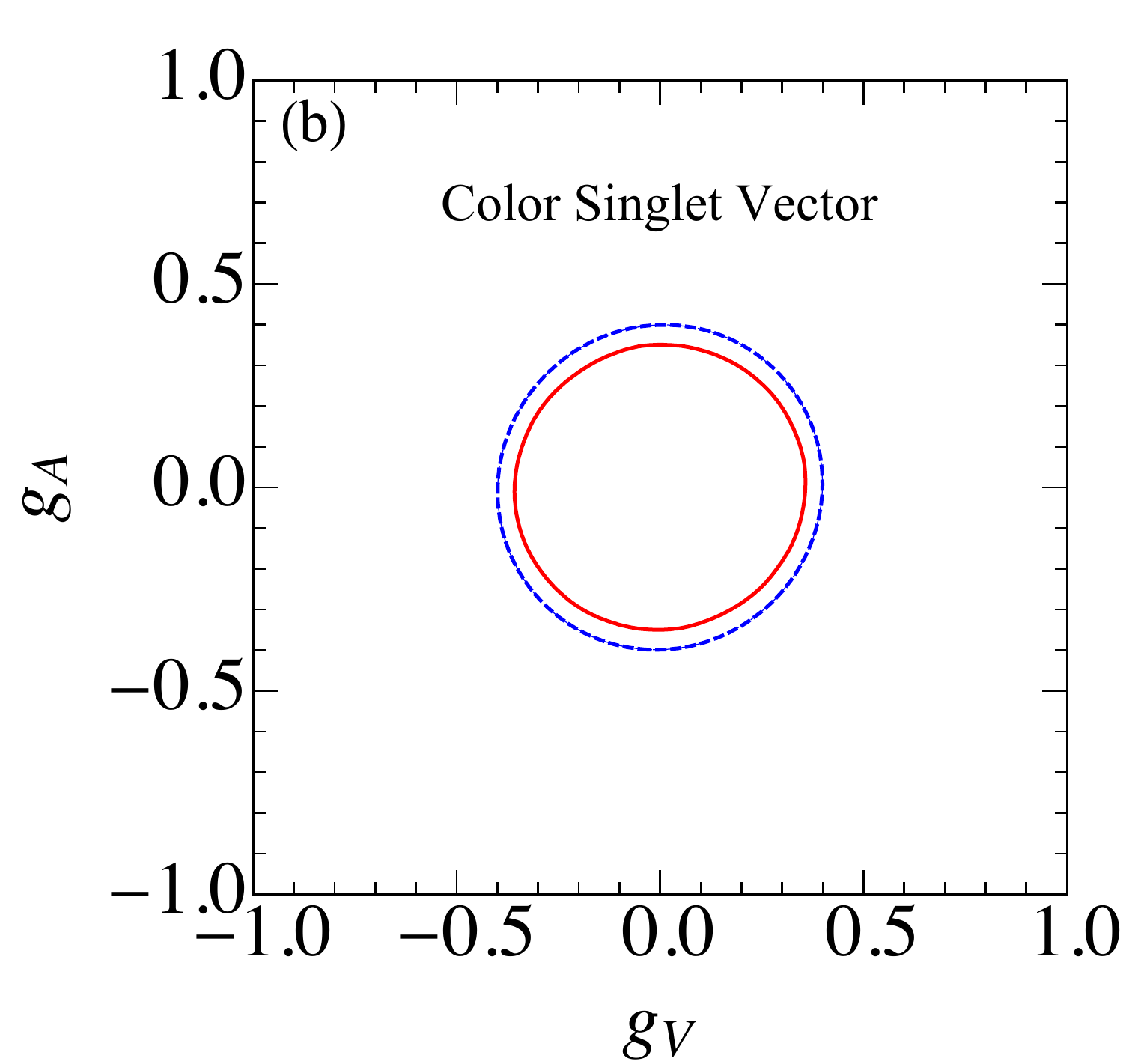}
\caption{Comparison between the full width (red) and the NWA (blue dashed): (a) color singlet scalar, (b) color singlet vector. For simplicity, the mass of the scalar and the vector are chosen both as 800 GeV.}
 \label{Fig:NWAcomp}
\end{figure}

Finally, we comment on the difference between the full width effect and NWA. As shown in Fig.~\ref{FIG:width}, for given mass and couplings, the width of a color neutral resonance is larger than a colored object, we focus on the color singlet scalar and vector. To emphasize the width effect, we choose a benchmark mass for both the scalar and vector as $M=800~{\rm GeV}$. Figure~\ref{Fig:NWAcomp} displays the allowed parameter space in the plane of effective couplings for a color neutral scalar (a) and vector (b), where the red solid curve represents the bound with the full width effects while the blue dashed curve the bound with the NWA. When the full width is incorporated into the calculation, the cross section of the $\tttt$ production is slightly enlarged to yield more stringent parameter space.

~\\
\noindent\textbf{3. High Dimensional Operators. }

In this section we consider two samples of high dimension operators which contribute to four top quark production. One is the dimension-6 four top-quark operator in composite top-quark models~\cite{Eichten:1983hw,Lillie:2007hd,Pomarol:2008bh,Kumar:2009vs,Zhou:2012dz} in which the right-handed top quark is composite. The four top quark contact operator is~\cite{Zhang:2017mls} 
\begin{align}
\mathcal{O}_{\tttt}=\frac{g_4}{\Lambda^2}(\overline{t}_R\gamma_\mu t_R)(\overline{t}_R\gamma^\mu t_R),
\end{align}
where $\Lambda$ denotes the NP scale. 
It yields the cross section of $\tttt$ production as following: 
\begin{align}
\label{EQ:high_dimensional_operator}
\sigma^{\rm total}= 9.608 - 1.637~g_4\left(\frac{\mathrm{TeV}}{\Lambda}\right)^2+ 4.664~g_4^2\left(\frac{\mathrm{TeV}}{\Lambda}\right)^4, 
\end{align}
where the constant term denotes the SM contribution, the linear term of $g_4$ is the interference between the SM and  $\mathcal{O}_{\tttt}$, and the quadratic term of $g_4$ represents the contribution from $\mathcal{O}_{\tttt}$ alone. The pure QCD corrections to the interference and quadratic term are calculated in \cite{Degrande:2020evl}, namely 0.57 and 0.93 respectively. We end up with a constraint on the $g_4$ coupling from the current data as 
\begin{align}
-1.34<g_4\left(\frac{\mathrm{TeV}}{\Lambda}\right)^2<1.55\,. 
\end{align}

\begin{figure}
\centering
\includegraphics[scale=0.5]{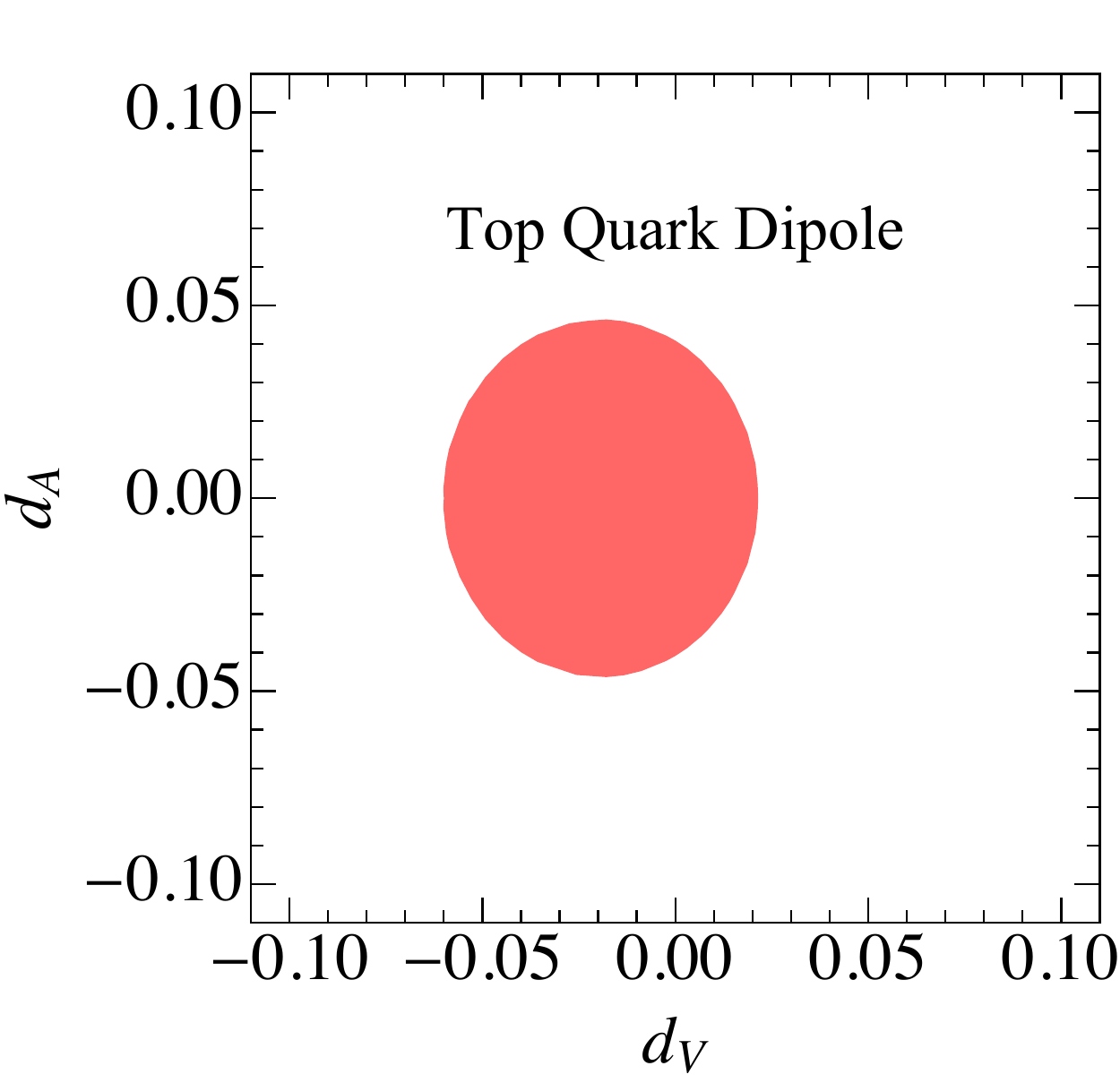}
\caption{\label{FIG:dipole} Allowed region for the Wilson coefficients of the top color electric dipole operator and color magnetic dipole operator. }
\end{figure}

The other sample is top quark dipole operator given by~\cite{Malekhosseini:2018fgp}
\begin{align}
\mathcal{L}\supset \frac{g_s}{m_t}\overline{t}T^A(d_V+id_A\gamma_5)i\sigma_{\mu\nu} t G^{\mu\nu A}. 
\end{align}
where $\sigma_{\mu\nu}\equiv[\gamma_\mu,\gamma_\nu]/2\,$. The dipole operator interferes with  the SM diagram in a complicated manner. For example,  its interference between the electric dipole operator and the SM depends mainly on the even power of the Wilson coefficient as the electric dipole operator violates the CP parity. It thus yields a symmetric bound on $d_A$.  On the other hand, the interference between the magnetic dipole operator and the SM depends on the odd power of Wilson coefficient and yields an asymmetric constraint on the $d_V$. Figure~\ref{FIG:dipole} displays the allowed parameter space in the plane of $d_V$ and $d_A$ with respect to the current LHC data. 

~\\
\noindent\textbf{4. A Broad Resonance. }

\begin{figure}
\centering
\includegraphics[scale=0.4]{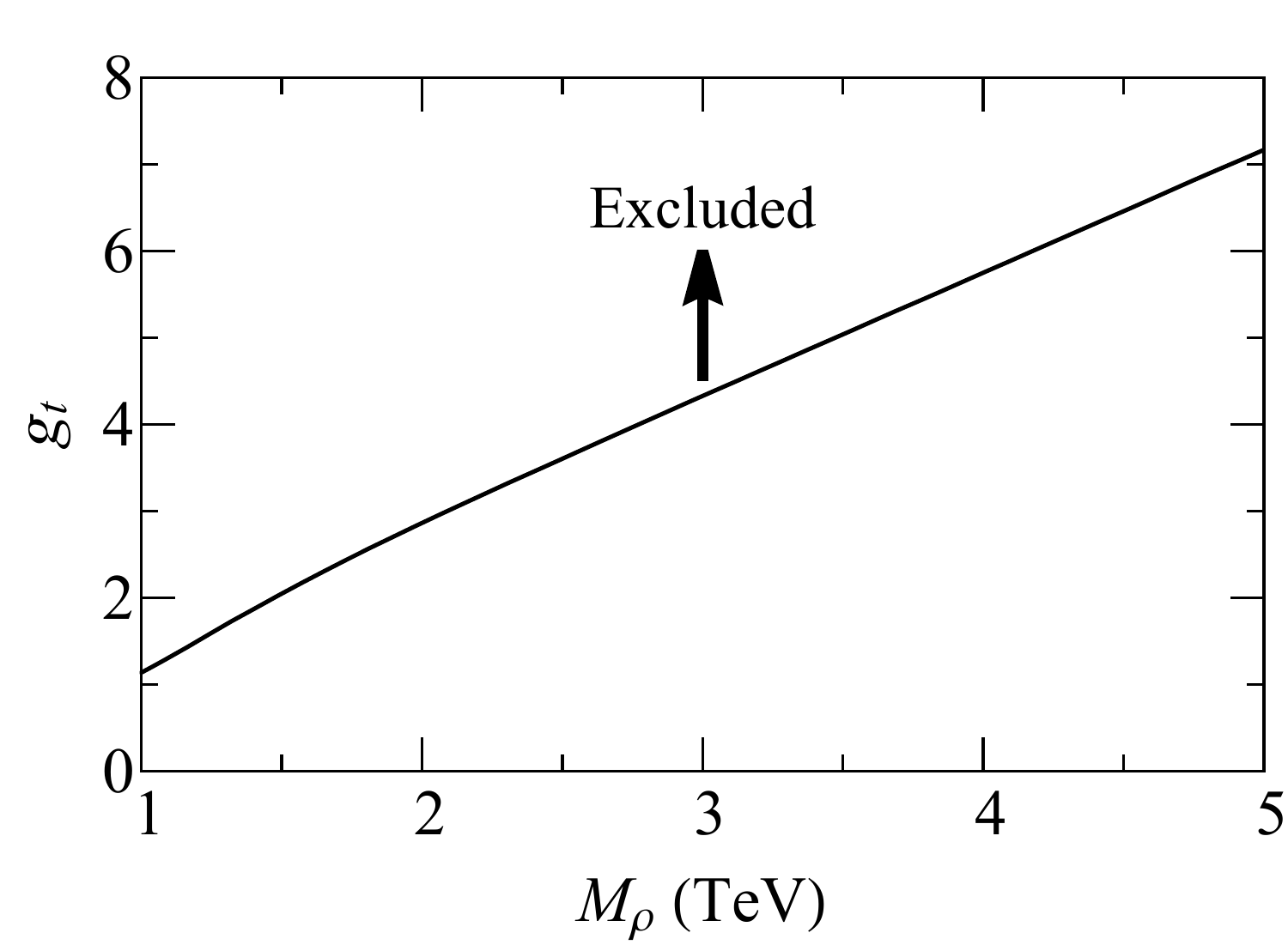}
\caption{\label{FIG:broad}The excluded parameter region for the broad vector resonance model. Only the large coupling region is shown. }
\end{figure}

At last, we investigate a special case of a broad vector resonance $\rho^\mu$ proposed in Ref.~\cite{Liu:2019bua}. The resonance decays mainly into a pair of top quarks or bottom quarks, and the Lagrangian reads
\begin{align}
\mathcal{L}\supset g_t(\overline{t}_L\gamma_\mu t_L-\overline{b}_L\gamma_\mu b_L)\rho^\mu\,,
\end{align}
where the coupling $g_t$ describes strong dynamics and can be fairly large. Such a large $g_t$ inevitably generates a very broad width of $\rho^\mu$. It is necessary to modify the Breit-Wigner distribution of the $\rho^\mu$ propagator to take care of the broad width.  
We modify the broad resonance propagator utilizing the CERN LEP line-shape scheme~\cite{Cao:2004yy}.
To be distinct from the previous study of the NP resonance with narrow width, we focus on the large $g_t$ and broad width here. Figure~\ref{FIG:broad} displays the exclusion limit of $g_t$ as a function of $M_\rho$, which shows that $g_t$ increases with $M_\rho$ linearly for a heavy $\rho$ with a mass of several TeVs. This is because the heavy resonance contribution to the $\tttt$ production depends on the quadratic and quartic powers of coupling-mass-ratio, e.g. $g_t/M_\rho$, after utilizing the modified propagator.

~\\
\noindent\textbf{5. Discussion and Conclusion. }

Despite the rare rate at the LHC, four-top production is the best channel to probe the so-called top-philic resonances that couple only to top quark. We studied various new physics resonances and, based on the current data at the LHC, explored the constraints on the mass of new resonance and the effective coupling of new resonance to the top quark. When the effective coupling is small, one can use narrow width approximation to simplify the study. However, when the coupling is large, the width effect is no longer negligible. A comparison of the full width effect and narrow width approximation is made in the study of the color-neutral scalar and vector. A special case of strong dynamics model is also addressed. 

~\\
\noindent\textbf{Note added:}
While finalizing the manuscript a nice work dealing with the same topic appears online~\cite{Darme:2021xxu}.

~\\
\noindent{\bf Acknowledgements}

The work is supported in part by the National Science Foundation of China under Grant Nos. 11725520, 11675002, 11635001, 11805013, 12075257, the Fundamental Research Funds for the Central Universities under Grant No. 2018NTST09, the funding from the Institute of High Energy Physics, Chinese Academy of Sciences (Y6515580U1) and the funding from Chinese Academy of Sciences (Y8291120K2).

\end{document}